\documentclass[12pt]{article}

 \newread\testifexists
 \def\GetIfExists #1 {\immediate\openin\testifexists=#1
     \ifeof\testifexists\immediate\closein\testifexists\else
     \immediate\closein\testifexists\input #1\fi}

 \usepackage{amsfonts}
 \usepackage{amssymb}
 \usepackage{graphicx} \usepackage{epstopdf}
 \immediate\openin\testifexists=e
     \ifeof\testifexists\immediate\closein\testifexists\else
     \immediate\closein\testifexists\input e\fipsf
 \def\Bbb#1{\setbox0=\hbox{$\tt #1$}  \copy0\kern-\wd0\kern .1em\copy0}
 \def\bbf#1{\setbox0=\hbox{$#1$} \kern-.025em\copy0\kern-\wd0
         \kern.05em\copy0\kern-\wd0 \kern-.025em\raise.0433em\box0}
 \immediate\openin\testifexists=a
     \ifeof\testifexists\immediate\closein\testifexists\else
     \immediate\closein\testifexists\input a\fimssym.def  % for \Bbb A - Z %%
 
 \newcommand{\tl}[1]{\tilde{#1}}              
                     \newcommand{\fn}{\footnote}
              
 \newcommand{\be}{\begin{eqnarray}}             \newcommand{\ee}{\end{eqnarray}}
 \newcommand{\bi}[1]{\begin{itemize}\item[#1]}           \newcommand{\ei}{\end{itemize}}
 \newcommand{\eqn}[1]{(\ref{#1})}

 \def\pa{\partial} \def\ra{\rightarrow}
  
 \def\dd{{\rm d}}

 \def\fract#1#2{{\textstyle{#1\over#2}}}
 \def\ffract#1#2{\raise .2 em\hbox{$\scriptstyle#1$}\kern-.3em/
                 \kern-.2em\lower .15 em \hbox{$\scriptstyle#2$}}
 
   \def\halff{\ffract12}

\def\bmatrix{\begin{matrix}} \def\ematrix{\end{matrix}} \def\bpmatrix{\begin{pmatrix}}\def\epmatrix{\end{pmatrix}}
\def\bcenter{\begin{center}} \def\ecenter{\end{center}}

\def\lowerheightfig#1#2#3{\(\raise-#1\hbox{\includegraphics[height=#2]{#3}}\)}
\def\lowerwidthfig#1#2#3{\(\raise-#1\hbox{\includegraphics[width=#2]{#3}}\)}

 \newcommand{\crlb}[1]{\label{#1}\\[2pt]}
 \newcommand{\eela}[1]{\quad\hbox{\scriptsize{#1}}\label{#1}\end{eqnarray}}
 \newcommand{\eelb}[1]{\label{#1}\end{eqnarray}}
 
 \newcommand{\newsecb}[2]{\section{#1}\label{#2}\setcounter{equation}{0}}

 \newcommand{\nolabels} {\def\eel{\eelb} \def\crl{\crlb} \def\newsecl{\newsecb}\def\bibiteml{\bibitem}\def\citel{\cite}}

\newcommand\publishversion{\nolabels\setlength{\textheight}{9in}\setlength{\oddsidemargin}{0in}
    \setlength{\textwidth}{6.3in}\setlength{\topmargin}{-0.1in}}

 \def\a{\alpha}      \def\b{\beta}   \def\g{\gamma}      
       \def\D{\Delta}  \def\e{\varepsilon} 
        \def\L{\Lambda}     \def\m{\mu}
                 \def\n{\nu}
                    
         \def\th{\theta}

\publishversion

     \def\OO{{\mathcal O}}  \def\NN{\mathcal{N}}

\begin{document} \begin{titlepage}

\title{\normalsize \hfill ITP-UU-15/04  \\ \hfill SPIN-15/02
\vskip 20mm{ \Large\bf  The Evolution of Quantum Field Theory}\\[10pt]
\large\bf From  $QED$  to Grand Unification}  

\author{Gerard 't~Hooft}
\date{\normalsize Institute for Theoretical Physics \\
Utrecht University \\ and
\medskip \\ Spinoza Institute \\ Postbox 80.195 \\ 3508 TD Utrecht, the Netherlands \smallskip \\
e-mail: {\tt g.thooft@uu.nl} \\ internet: {\tt
http://www.staff.science.uu.nl/\~{}hooft101/ }
\vfill
A contribution to:\\
The Standard Theory up to the Higgs discovery\\
--- 60 years of CERN ---\\
L.~Maiani and G.~Rolandi, eds.
}

\maketitle

\end{titlepage}
\eject
% \setcounter{page}{0}
%%%%%%%%%%%%%%%%%%%%%%%%%%%%%%%%%%%%%%%%%%%%%%%%%%%%
%\def\E{\hbox{\scriptsize{E}}}

\begin{quotation} \noindent {\large\bf Abstract } \medskip \\
	In the early 1970s, after a slow start, and lots of hurdles, Quantum Field Theory emerged as the superior doctrine for understanding the interactions between relativistic sub-atomic particles. After the conditions for a relativistic field theoretical model to be renormalizable were established, there were two other developments that quickly accelerated acceptance of this approach: first the Brout-Englert-Higgs mechanism, and then asymptotic freedom. Together, these gave us a complete understanding of the perturbative sector of the theory, enough to give us a detailed picture of what is now usually called the Standard Model. Crucial for this understanding were  the strong indications  and encouragements  provided by numerous experimental findings.  Subsequently, non-perturbative features of the quantum field theories were addressed, and the first proposals for completely unified quantum field theories were launched. Since  the use of continuous symmetries of all sorts, together with other topics of advanced mathematics,  were recognised to be of crucial importance, many new predictions were pointed out, such as the Higgs particle, supersymmetry and baryon number violation. There are still many challenges ahead.
	\end{quotation}
{\vfill \flushleft{March 12, 2015}}\\

\setcounter{page}2
\setcounter{section}0

\newsecl{The Early Days, before 1970}{early}
Before 1970, the particle physics community was (unequally) divided concerning the relevance of quantized fields for the understanding of subatomic particles and their interactions. On hindsight, one can see clearly why the experts were negative about this approach.  Foremost was the general feeling that this theory was ugly, requiring various fixes to cover up its internal mathematical inconsistencies. The first inconsistency, as it was generally perceived, was the fact that the corrections to the particle interaction properties, generated by higher order quantum effects, invariably appeared to be infinitely strong. The energy contents of a field surrounding a particle, would clearly add an infinitely large correction to its mass, and also electric charge and other interaction parameters would receive infinite corrections by vacuum fluctuations in the vicinity of a particle.

	Now, it was true that a remedy had been proposed to this particular disease, first perhaps by Hans Kramers\,\cite{kramers} around 1933, and more precisely by Julian Schwinger\,\cite{schwinger}, Freeman Dyson\,\cite{dyson}, Sin-Itiro Tomonaga\,\cite{tomonaga}, Richard Feynman\,\cite{feynman1} and others, which was that the `original' masses and interaction strengths of a particle are ill-defined, so that these could be adjusted to cancel out the unwanted infinities, which were now replaced to experimentally inaccessible regions near the cores of these particles. A systematic application of this procedure, called \emph{renormalization}, turned out to be quite successful in the study of electromagnetic forces between particles\,\cite{mehra}. The anomalous magnetic moment of the electron thus obtained agreed extremely well with experimental determinations, and other successes of this theory, called \emph{Quantum Electrodynamics} (QED), soon followed\,\cite{g-2}.
	
	Yet there were reasons to mistrust these results. The mathematical rigour of the calculations was lacking, it looked as if the difficulties had been swept under the rug. Perhaps these arguments were approximately right for QED, but what were the principles lying behind the other interactions? And how can we understand the renormalization procedure from a more formal point of view? Indeed, if one attempted to understand the small-distance limit of QED, a new difficulty showed up: the interactions due to virtual particles in the surrounding vacuum accumulate there, and in spite of renormalization, the effective interaction strengths eventually tend to infinity. Today, we know that that is because, in the small distance limit, you are looking again at the bare masses and coupling strengths, and these had just been agreed upon to be infinite. Lev Landau saw that this infinity would behave like a physically unacceptable `ghost particle', now called `Landau ghost'\,\cite{landau}. Today, we know how to handle the Landau ghost, but for Landau this clearly implied that you had to abandon quantized field theories altogether.
	
	In the West, investigators were a bit bolder. Murray Gell-Mann an Francis Low had proposed that there could be an ultraviolet fixed point\,\cite{gellmannlow}. That, however, did not help very much because this fixed point would be in a domain where accurate calculations are impossible. It looked as if Nature was telling us that the real particle spectrum is more subtle, and to understand that, you will have to start from scratch. Stay away from quantum field theory.
	
	Indeed, experimental results were not encouraging at that time. The weak force appeared to be non renormalizable for simple mathematical reasons (Enrico Fermi's fundamental interaction constant carried the wrong dimension if you simply considered how it was defined). If you would try to replace Fermi's original theory by a theory of exchanged intermediate particles (now known as the \(W\) bosons), you would end up with fundamental spin-one force carriers, particles that appeared to require a totally different approach as well. And, most of all, the strong force seemed to resist any rational approach altogether. The hadron spectrum suggested the existence of sub-units called "quarks" by Gell-Mann\,\cite{gm8}, and `aces' by George Zweig\fn{Gell-Mann thought of three fundamental quarks, but Zweig, as he would explain later, assumed that there should be four, thus anticipating the idea of charm. There are four aces in a deck of cards}\,\cite{zweig}, but how a field theory could be responsible for that strange situation was beyond us.
	
	It is always dangerous to combine several non convincing arguments to reach a conclusion with certainty, but this is what almost happened. Quantum field theory was not \emph{bon ton}. Indeed, there were alternatives. You could start at a more basic level. When an experiment with subatomic particles is carried out, one begins with beams of particles directed towards one another, the so-call \emph{in-} situation, or in-state. After the particles collided, you end up with different beams of particles going \emph{out}, the out-state. The out-state depends on the in-state chosen. This dependence is described by a matrix called \(S\)-matrix ($S$ for \emph{Streuung}, or \emph{scattering})\,\cite{smatrixheisenberg}. One can derive mathematical equations that this \(S\)-matrix must obey. By demanding that no signal can ever travel faster than light, one finds the so-called \emph{dispersion relations}, relations between frequencies and  wavelengths\,\cite{vank}, and more general features in multi-particle scattering.
	
	It was hoped that the \(S\)-matrix could be derived from such relations, if combined with some general symmetry properties. To this end, \emph{current algebras} were constructed\,\cite{chewcurrent}. What investigators tried to avoid is to talk of operators acting at single space-time points. That would have been helpful in the current algebras, but it smelled too much like field theory.
	
	Searching for totally different principles, Steven Frautschi and Tullio Regge\,\cite{reggefrautschi} attempted to consider amplitudes as functions of angular momentum, which could be analytically continued to the complex plane. This yielded the famous `Regge trajectories', curves that could be extended to include all resonances, giving useful but ill-understood relations between spin and energy.
	
	Only a handful of researchers resisted the mainline thought. First of all, QED was further refined, and appeared to work beautifully\,\cite{feynman2}. Quantities such as the anomalous magnetic moment \(g-2\) of the electron and the muon, could be calculated and compared with experiment up to an incredible precision. It would be great if anything like that could be constructed to describe any of the other forces. 
	
	\emph{Suppose we forget those negative preconceptions about field theory, forget even the experimental data, and instead, just ask the question: what shape could a fundamental interaction possibly have?} 
	
	Maybe an axiomatic approach would help. We had Arthur Wightman's famous axioms\,\cite{wightman}, idealising the demands any quantized field theory should obey. On hindsight, one may say that these demands were too strict; even today's Standard Model formally does not obey them, but as its breakdown occurs somewhere beyond the Planck scale, no-one cares about that anymore.
	
	In a more practical vein, a `toy model' had been coined by Gell-Mann and Maurice L\'evy, to describe strongly interacting pions in agreement with their symmetry structure, a spontaneously broken global symmetry called \emph{chiral symmetry}. At that time, this was phrased in terms of a ``partly conserved axial vector current" (PCAC)\,\cite{gmlevy}. The model worked qualitatively well, certainly in connection with the famous Goldberger-Treiman relation\,\cite{goldbtr}, but, being an ill-understood strong interaction theory, it could not be expected to be very accurate. Ingenious resummation techniques were attempted, but such attempts, as would also be demonstrated at several occasions later, are fruitless if one does not understand the underlying physics.
		
	And then there was the Yang-Mills theory\,\cite{ym}. What Frank (C.N.) Yang and Robert Mills had done, way back in 1954, would turn out to be extremely important: \emph{} Indeed, if the only two force theories that are really successful, being electrodynamics and Einstein's General Relativity, are both based on some fundamental local symmetry, are there other ways to employ symmetries in a similar way, to describe different forces? When I was M.~Veltman's undergraduate student, he would already point out this paper to us. ``This you must know", he said, ``this is very important". When I asked why, he said, ``I don't know, just read it."

	But the theory they came up with seemed to make no sense. Yang-Mills theory required the existence of massless spin-one particles, much like photons, except that, unlike photons, they carry charges themselves. Such particles were known not to exist, that is, if charged spin-one particles exist, they must have mass. Yang and Mills were wise enough nevertheless to publish their result. That result was a new kind of quantum field theory.

\emph{Without understanding the physics, mathematics does not answer your questions. Without understanding the mathematics, your physics theories will not work successfully either}, is what we had to discover (and we keep forgetting time and again).

	What Veltman had seen, was that there seemed to be a deep connection between the experimental data concerning the weak force, and Yang-Mills theory.  Martin (Tini) Veltman\,\cite{MV}  commenced his own personal battle to make sense out of these strange observations. This was a quantum field theory, it had infinities that had to be renormalized, and Nature appeared to be telling him that these ideas should work. Nobody really understood the physics, but he did understand which mathematical equations had to apply. These were so complicated that he decided to construct a computer program to address lengthy equations. ``Schoonschip", was the name of his program, a word that only Dutch citizens can pronounce, so that his property rights would be guaranteed. Schoonschip told Veltman that, indeed, there was something wrong with the physics of the Yang-Mills theory.
	
	An other obscure corner was investigated by Peter Higgs\,\cite{higgs}, Robert Brout and Fran\c cois Englert\,\cite{be}. They enjoyed little attention when they argued that the symmetry employed by Yang and Mills had to be spontaneously broken. The reason for that was that this alley had also been closed by the ``experts". There was the famous `Goldstone Theorem'\,\cite{goldstone}: \emph{Whenever a symmetry is spontaneously broken, at least one particle must become massless}. Indeed, in the Gell-Mann L\'evy Model, the pion behaves as a massless particle. The weak interaction, however, did not seem to involve massless objects. Higgs, Brout and Englert saw no massless particles in their models either, but a major fraction of the community did not believe them. So they were mainly being ignored. Veltman paid no attention at all to formal mathematics, so he believed neither Higgs, Brout and Englert, nor Jeffrey Goldstone. He only believed the experiments, and his computer.
	
	It was in this atmosphere that, independently, three people did foresee models for the electric and weak forces that would later turn out to be the precursors of the Standard Model. Abdus Salam\,\cite{salam} gave some general talks advocating theories resembling what is now called the BEH mechanism to understand these forces. Shelley~Glashow\,\cite{glashow} saw how Yang-Mills photons, slightly modified to make them massive, could generate quite neatly weak forces as observed in the experiments, and Steven Weinberg\,\cite{weinberg} wrote down the most detailed theory for the entire lepton sector\fn{Weinberg left out all hadronic weak interactions, and this was for a very good reason: the hadrons did not seem to fit in his model. Weinberg understood that his model would predict strangeness changing neutral current interactions, while these were not observed in the experiments. The GIM mechanism, only to be discovered later, would turn out to be the explanation of this apparent contradiction.} , including the effects of the Higgs particle. They were mostly ignored, and even the authors themselves continued working on other subjects. The unsolved problem was how to renormalize these theories.
	
	\newsecl{The new ideas of the 1970s}{1970}
	
	Historians often talk of a `crisis' that precedes one or more revolutions of thought for the realisation of new breakthroughs. I don't think that applies here. There was no crisis, new experimental results were coming in, the nature of our problems was clearly identified, and there were plentiful ideas. Yet, we had no advance warning that new landslides were ahead, and these came, in a very quick succession. Problems that at one time had looked hopelessly complex, were solved with unexpected elegancy, and when the clouds lifted, we had a beautiful and relatively simple ``Standard Model" for all known subatomic particles.
	
	It is also not true that our work on Yang-Mills theories was motivated by our wish to put the Standard Model on a proper mathematical footing, as the story is now often told. The Standard Model wasn't there yet, the only existing theories that had a more or less proper mathematical footing were QED, and models that include purely scalar fields, which did not seem to apply to anything. Landau's difficulty was still there, in both these systems. We wanted to understand how to deal with infinities when you have fundamental vector particles (particles with spin one). Veltman had seen correctly that the infinities are particularly mild when the interactions have a Yang-Mills structure. Two more things had to be done.
	
	First, we needed to understand how the original Yang-Mills theory would have to be treated as a quantum field theory, and how its interactions have to be renormalized, without jeopardising the local symmetry structure. This meant that one cannot simply say that \(\infty\ -\ \infty\ =\) something finite, but one has to establish how these finite expressions reflect the correct symmetry structure and dispersion relations.\fn{Indeed, the old ideas to use dispersion relations and symmetries were still quite useful!} The local symmetry is a gauge symmetry, just as in electro magnetism, and so, these theories were also called gauge theories.
	
	What are the Feynman rules? Feynman had discovered that the mathematical equations for field theories can be framed in terms of neat sets of rules. In the new theories, however, Feynman's rules could be phrased in many ways, and they did not seem to be equivalent. The original idea was that the propagation of particles (real or virtual ones) is represented by lines, called `propagators',  in Feynman's diagrams, but now we seemed to get lines that do not describe a particle at all. Worse even, early investigations by Bryce DeWitt\,\cite{DeWitt}, Ludwig Faddeev, Victor Popov\,\cite{faddeev}, Richard Feynman\,\cite{feynman3}  and Stanley Mandelstam\,\cite{mandel} all seemed to produce \emph{different} rules for the propagators! 
	
	We had to understand what the new rules for these `ghost particles' would be. 
This problem was solved\,\cite{GtH-1}, the hard way, meaning that we analysed diagram by diagram. We found that there are many different ways to generate Feynman rules, so that DeWitt, Faddeev, Popov and Mandelstam all were using correct rules, except Feynman himself: he had used the rules for \emph{massive} Yang-Mills theories, which are not the same, because there is a third spin direction that does not go away when you send the mass to zero. This is why Feynman was unable to ge beyond one loop: he used the wrong theory (massive Yang-Mills theory without a physically observable Higgs particle, was later found not to be reormalizable).

	To see how unitarity and dispersion relations work out in a gauge theory, we selected the proper equations that should be obeyed by the renormalized diagrams. These equations were the non-Abelian generalisations of the older Ward-Takahashi identities for QED, which we needed only when the external lines are on their mass shells. The identities looked like symmetry relations, but we could not identify the symmetry, because of all sorts of curious minus signs everywhere. 
	We did not bother to work out how our relations have to be modified when we go off mass shell, although our combinatorial proofs did use off-shell diagrams. This omission was quickly corrected when, independently, Andrei Slavnov\,\cite{slavnov} and John C.~Taylor\,\cite{taylor} wrote down the more complete expressions.

Although we thought that our proofs worked just fine, not everyone was happy with our diagrammatic formalism. It was a few years later when Carlo Becchi, Alain Rouet, Raymond Stora and Igor Tyutin\,\cite{brst} found a more elegant procedure to handle the Feynman rules: they discovered a curious, apparently unphysical supersymmetry, now called BRST symmetry, which holds for theories where gauge conditions have been chosen, while they ensure the possibility to transform to different gauge choices. BRST made extensive use of the Slavnov-Taylor identities. So these \emph{were} symmetry relations, and our curious minus signs were trying to tell us that this was a supersymmetry!

With these problems out of the way, we seemed to be ready to renormalize the theory. All one has to keep in mind that the BRST symmetry should not be disturbed. From where we were now, it was also not difficult to give mass to the Yang-Mills bosons\,\cite{GtH-2}. 	The next step that we planned to take, turned out to have been analysed earlier: the older papers by Englert, Brout, Higgs and Weinberg\,\cite{higgs}\,\cite{be}\,\cite{weinberg} were quickly unearthed and found to be relevant. The result of our mathematical excursion could almost have been guessed (as Salam and Weinberg had done): \emph{If you write the Lagrangian density for the theory, you read off all dynamical variables and all interaction parameters. They must all have a strict canonical form; in that case, a systematic perturbation expansion can be set up, and if the interaction parameters are not too big, you get a very accurate theory.} 	
	
	However, an other difficulty showed up: anomalies. It could easily happen that when a symmetry property is imposed on one aspect of an interaction amplitude, a violation of a similar symmetry property elsewhere pops up, as had been noted by Steve Adler, John Bell and Roman Jackiw\,\cite{abj}. These anomalies resemble a lid that does not properly fit onto a jar. One such anomaly causes neutral pions to decay into two photons, while chiral symmetry would have forbidden such a decay. Do we have non-Abelian gauge anomalies? If so, these would be standing in the way of our renormalization procedure.
	
	The problem would be solved if we could find a `gauge invariant regulator'. A regulator is some procedure, invoking some hidden physical phenomenon, that makes the theory finite. We wanted such a regulator that respects local gauge symmetry. We knew the regulator that was often used for QED. It had been found by Wolfgang Pauli and Felix Villars\,\cite{pv} that one can introduce `very heavy ghost fermions'  that do the desired trick. We did not succeed in finding such ghost particles that work in the Yang Mills case. One trick was a procedure that would later become popular as the `Kaluza Klein theory': Theodor Kaluza and Oskar Klein\,\cite{kk} had proposed to employ an extra `fifth dimension'. 
Particles moving in the fifth dimension could act as our regulator, but only for the first quantum corrections, the ones described by Feynman diagrams with only one closed loop in them. In spite of vigorous attempts, we could not tame the diagrams with more than one, overlapping, loops. 6 or 7 dimensions perhaps? To no avail.
	
	The answer turned out to be that one has to use a continuously varying number \(n\) of dimensions, choose \(n=4+\e\), where \(\e\) is only infinitesimally small, and take the limit \(\e\ra0\) sufficiently carefully\,\cite{GtHMV}. That worked! This method, to be called `dimensional regularization and renormalization, would later turn out to have the additional advantage that the algebra stays really simple, so that calculations can be done quickly and efficiently. This way we learned how to renormalize the pure Yang-Mills theory up to all orders in the quantum corrections. This theory would now be at least as good as QED! 
		
It soon turned out that dimensional regularisation had been discovered independently and practically simultaneously by C.G.~Bollini and J.J.~Giambiagi\,\cite{bollgiam} and also by J.F.~Ashmore\,\cite{ashmore}. They applied it to QED, showing that gauge-invariance is maintained there.
	
	However, the anomalies give an extra twist to the story: the axial vector current is special to a theory in 4 space-time dimensions, so that dimensional renormalization does not directly work for that kind of current. This is why anomalies can occur in its conservation law. If gauge fields couple to axial vector currents, one must check explicitly whether these currents are properly conserved. \emph{The anomalies must all cancel out, which requires a special relation between left helicity and right helicity fermionic particles.} Quite generally, this check needs to be performed only for the 1-loop diagrams, as was already understood by Adler and Bardeen\,\cite{adlerbardeen}. If the chiral anomaly cancels out here, it will cancel out at all higher loop levels as well.
	
	The importance of the discovery how to renormalize Yang-Mills theories with BEH mechanism, was almost immediately realised by a majority in the particle community. Benjamin W.~Lee, Kurt Symanzik, Jean-Loup Gervais and Pronob Mitter had been lecturers in the 1970 Carg\`ese School of Subnuclear Physics, discussing the Gell-Mann L\'evy sigma model, so that they knew the importance of spontaneous symmetry breaking. Sidney Coleman, as a guru of mathematical physics at the time, embraced the new and important role of mathematical group theory. Also, all were delighted to see that now the rules of the game had been made clear. One can write down models, generalisations of, or alternatives to, the Standard Model, and immediately read off their main predictions.
	
	One of the predictions was due to the practically unavoidable presence of a neutral component of the carriers of the weak force. Now, the effects of this \emph{neutral current interaction} could be precisely calculated. It was predicted that the behaviour of neutrinos would be affected,  as now they can scatter elastically against electrons, which could be confirmed with ingenious experiments\,\cite{haidt}. Also, it was now strongly suspected that the quark spectrum known at the time could not be complete. The \emph{charmed quark}, proposed by Glashow, John Iliopoulos and Luciano Maiani\,\cite{gim}  in 1970   was needed to accommodate for the left-right asymmetry in the weak interactions; it explained the absence of flavor changing neutral current effects, and it was also needed to cancel out the chiral anomaly there.
	
	Why were all these findings so remarkable? We now know that the resulting model could not be \emph{infinitely} accurate. The canonical conditions on the parameters of the theory just happen to guarantee that perturbation expansions can be carried out \emph{up to any order of the expansion}. The only thing that can go wrong -- and it does -- is that the expansion will not converge. In practice, however, this is only a formal difficulty;  one can calculate the consequences of the theory much more precisely, in principle, than any measurement -- provided the couplings are not too strong. So, paradoxically, the successes of the renormalization program for perturbative Yang-Mills theories are due to the fact that our theory cannot be the ultimate theory of Nature\fn{Here, we talk of the \emph{purely theoretical} argument that the Standard Model is not infinitely precise. Besides that, we would later encounter other evidence from experimental observations (dark matter) and phenomenology (the failure of exact grand unification) that the Standard Model is incomplete.}.
	
\newsecl{The strong interactions}{strong}
	
	In these early days, there were other, more urgent difficulties . The \emph{strong} interactions of course require strong interaction parameters, so by their very nature, these would not allow for a perturbative treatment. We had the important suggestion, inspired by experimental observations, and the ensuing phenomenological theories, that hadrons are formed of quarks, but these were never observed. It seemed that a fundamental new principle was at work here. 
	
	The resolution to this difficulty came again quickly and unexpectedly. Some mathematical features of the renormalization procedure had been investigated, way back in the 1950s, by Stueckelberg and Petermann\,\cite{peterm}. The freedom of choosing which part of the interaction parameters should be used to cancel infinities at higher orders, the core of the renormalization program, should manifest itself in a certain group property of the theory, which they called the `renormalization group'. Actually, only that part of the renormalization group that gets involved in \emph{scale transformations} lead to novel features that would otherwise be difficult to understand. So it happened that the renormalization group became practically synonymous to the group of scale transformations.
	
	Invariance under renormalization group transformations was cast into equations by Curtis Callan\,\cite{callan} and Kurt Symanzik\,\cite{symanzik}. These equations contain new functions \(\a\), \(\b\), and \(\g\), all depending on the coupling parameters, and describing what happens under scaling. Of these, particularly the function \(\b(\vec g\,)\) became important, where \(\vec g\) stands short for the set of coupling parameters\fn{There was also a, less significant, dependence on mass parameters \(\vec m\).}. \(\b\) tells us where the Landau ghost is. The crucial thing to find out is its sign.
	
	 Theorists thought they knew everything about that sign. The sign is positive. At least, this is so for QED and the theories with scalars, and it was almost proven to be a universal fact. This would mean that \emph{every} quantum field theory should have its Landau ghosts, and these would cripple the theory. Moreover, if there are strong interactions, the Landau ghosts would be close by and nasty. In fact, this had been Landau's reason to dismiss quantum field theories altogether.
	 
	 The story of the sign of the \(\b\) (beta) function is one full of misunderstandings and miscommunications. How do vector particles (particles with spin 1) contribute to beta? How can renormalization counter terms switch sign? Why did everyone believe that the sign had to be positive?
	
	The latter mistake is actually quite understandable. The renormalization of charge-like parameters is due to clouds of virtual particles hovering near a charged particle, due to vacuum fluctuations. A spinning particle is just like a spin-less particle of which there are several species. Why should their effects depend so much on spin at all? Why should the \emph{sign} change due to spin? Rough estimates, in stead of accurate calculations, would indeed suggest a universal sign. Around 1970,  David Gross was firmly committed to prove the universal sign of the beta functions, and he was about to publish his proof\,\cite{gross1}.
	
	In 1965, Vladimir Vanyashin and Mikhail Terentyev\,\cite{vt} found a negative sign in the charge renormalization of charged vector mesons, but they attributed this `absurd' result to the non renormalizability of the theory. In 1969, Iosif Khriplovich\,\cite{khripl} correctly calculated the charge renormalization of Yang-Mills theories in the Coulomb gauge, where there are no ghosts. He found the unusual sign, but his important result was not noticed.
	
	The story of the calculations of the \(\b\) function is given in more detail by Misha Shifman\,\cite{shifman}. He concludes  that asymptotic freedom was not noticed before 1973, when David Politzer, David Gross, and Frank Wilczek published their results\,\cite{politzer}\cite{grosswilczek}, but actually the story told says something altogether different: asymptotic freedom was discovered three times\fn{I am sure of the third case.} before 1973, but not recognised as a new discovery. This is just one of those cases of miscommunication. The ``experts" were so sure that asymptotic freedom was impossible, that signals to the contrary were not heard, let alone believed. In turn, when I did the calculation I found it difficult to believe that the result was still not known.
		
	 In the mean time, in a very different topic of research, James (BJ) Bjorken\,\cite{bjorken} had found that scaling properties of hadrons may be easy to explain if, at very high energies, constituent particles of hadrons are weakly interacting. This was called Bjorken scaling. Now this would require a \(\b\) function that is negative or vanishing. It was suspected that Bjorken scaling should therefore be the strongest argument \emph{against} quantum field theories.
	
	The author had done his calculations on how Yang-Mills theories scale in 1971\fn{A brief remark in Ref.\,\cite{GtH-2}, at the bottom of the first page,  refers to this result, which he  expected to be known to the other investigators.}, found the negative sign, and decided he understood nothing of the arguments people had against field theories for the strong interactions. He did not see any Landau ghost there, and started to investigate strong interactions his own way. The obvious candidate was a pure \(SU(3)\) Yang Mills theory \emph{without} BEH mechanism. Now how does one \emph{prove} that the quarks do not emerge as free particles? To advance such a theory, one would have to come with a credible explanation of the \emph{quark confinement mechanism}. I was told that I should not make a fool of myself by publishing a theory of the strong force if \emph{that} problem could not be addressed. This held me up. In June 1972,  I did communicate my result about the negative \(\b\) function, verbally at a conference in Marseille. One of the attendants was Symanzik, who strongly urged me to publish it. ``If you don't, someone else will", he said, and right he was. 
	
	Politzer\,\cite{politzer}, a student of Coleman's, and independently Gross and Wilczek\,\cite{grosswilczek}, were the first to publish their finding that \(\b(g)\) is negative for Yang-Mills fields, in 1973. They also understood its significance for understanding Bjorken scaling, and how it could help understanding quark confinement (in a qualitative way). This was the beginning of a more precise understanding of the strong force. The basis for the new theory had already been proposed by Harald Fritzsch, Murray Gell-Mann and Heinrich Leutwyler\,\cite{fritzsch} in 1971, but in their earliest ideas they still had to struggle with the confinement problem, and the problem of the high-energy behaviour -- the Landau ghost still seemed to be there. They also coined the new name for this theory: \emph{Quantum Chromodynamics} (QCD).
	
\newsecl{The first years of the Standard Model. Quantum Chromodynamics.}{earlySM} \def\left{{\mathrm{left}}} \def\right{{\mathrm{right}}}

By now, we had some glimpses of a new synthesis. In total, the local gauge symmetry structure appeared to be \(SU(3)_{\mathrm{strong}}\otimes SU(2)_{\mathrm{weak}}\otimes U(1)_{\mathrm{EM}}\). The leptons known at the time came in two \(2_\left\oplus 1_\right\) representations, the quarks in two \(3\otimes 2_\left\oplus (3\oplus 3)_\right\) representations (the third family would come somewhat later). The BEH mechanism involved a complex scalar doublet field. Three of its four real components provide mass to the charged \(W\) bosons and the \(Z\) boson. One component, the radial one, is gauge-invariant and should have observable energy quanta: the Higgs particle. It was known that the model we had here was not yet complete; there was no mechanism yet for \(CP\) violation, and it was not known whether neutrinos had mass. If they do, we would need extra neutral (``sterile") fermionic fields for them. It had been proposed by Makoto Kobayashi and Toshihide Maskawa\,\cite{km} that a third family of quarks and leptons would provide for a natural mechanism for \(CP\) violation; the various members of this third family were discovered over the decades that followed.

An urgent theoretical problem was the better understanding of QCD. Just before the gauge theory revolution another development had taken place: the dual resonance models. Gabriele Veneziano's phenomenological expression for the elastic scattering of two mesons, yielding realistic descriptions of sequences of higher spin resonances, had been generalised to encompass the \(n\)-particle amplitude by Ziro Koba and Holger-Bech Nielsen\,\cite{kobaniels}. They, and Yoichiro Nambu and Lenny Susskind, realised how these expressions could be interpreted physically: these mesons are pieces of a quantized, relativistic string. 

Now this could be reconciled with the quark picture of QCD if the forces holding quarks together can be described in terms of narrow vortex lines connecting the quarks. If these vortices may also be assumed to expand somewhat in the two transverse dimensions, then a clear picture arises of the behaviour of the QCD-gluon fields: they condense into vortex structures. The question was then, how to understand this behaviour starting from the Yang-Mills Lagrangian for QCD.

Here, an insightful observation by Nielsen and Poul Olesen\,\cite{nielsols} was very helpful. If one considers an Abelian Higgs theory, then this Higgs field is a single complex scalar. Such a field allows for a topologically stable string-like configuration, the Nielsen-Olesen vortex: this vortex occurs whenever the complex Higgs field makes a full rotation in the complex plane. This feature is well-known in the BCS theory for superconductors. A material that is infinitely conducting cannot contain a magnetic field; magnetic fields are shielded. If however, a magnetic field gets stronger than some limit, the superconductor becomes unstable, temporarily looses its superconductivity, and is forced to allow the magnetic field in. One then finds that such a field takes the shape of a vortex, the Abrikosov vortex, the solid-state analog of  the Nielsen-Olesen vortex.

The Nielsen-Olesen vortex, and the Abrikosov vortex, should be stable, unless they can break in pieces; in the letter case, one should be able to describe what an end point looks like. In superconducting material, this is clear: the vortex carries magnetic flux, so end points can only occur if we have magnetic monopoles. In ordinary physical systems, magnetic monopoles do not exist, and so the Nielsen-Olesen vortex is stable. Incidentally, since the total magnetic flux in a vortex must be quantized in units of \(2\pi/e\), where \(e\) is the electric charge quantum, this is a simple way to see that also the magnetic charge \(g_m\) of magnetic monopoles must be quantized the same way. This was already known by Paul Dirac\,\cite{diracmonopole}.

Curiously, Julian Schwinger\,\cite{schwingermonopole} had arrived at \(4\pi/e\) as the unit of magnetic charge. He had a problem with Dirac's value, which is half of that, so he thought Dirac's value is impossible.  We'll show in Section~\ref{mm} that indeed something special is going on at Dirac's charge, but our theories do no forbid that.

Can Nielsen-Olesen vortices also help understanding permanent quark confinement? At first sight, no. This would require quarks to behave as magnetic monopoles, but these were difficult to describe. If, as stipulated by Bjorken, at very high energies quarks behave nearly as free particles,  their magnetic monopole charges, having values such as \(2\pi/e\), should be weak, so that the gauge field couplings \(e\) should be very strong -- this would be a contradiction.

What would be the non-Abelian analogue of the Nielsen-Olesen vortex? Here, something interesting happens. The stability of the vortex hinges on the question whether a closed loop in the space of gauge transformations is contractible or not. If it is contractible, we say then that the gauge group is simply connected, the vortex is unstable, otherwise it is stable. In mathematical terms, the quantisation of vortices is controlled by the elements of the homotopy group \(\pi_1\) of the gauge group \(G\). Now, suppose we have a Higgs mechanism, breaking a gauge group such as \(SO(3)\) into a subgroup, say \(SO(2)\), or equivalently, \(U(1)\). For \(U(1)\), the homotopy group is \(\pi_1(U(1))=\Bbb Z\), the addition group of the integers. For \(SO(2)\), however, \(\pi_1\) is only \(\Bbb Z(2)\), the group of additions \emph{modulo} 2. This means that if we would add two vortices, they can annihilate each other. In terms of magnetic monopoles, if you combine two monopoles with the same magnetic charge, they could annihilate one another.

This implied that, after the BEH mechanism is switched on, the fields could carry away a magnetic monopole charge to an amount of \(4\pi/e\), or: it should be possible to construct a magnetic monopole with charge \(4\pi/e\) out of the gauge fields in this model!

Once we realised that the gauge fields in a BEH system can generate magnetic monopoles, it was not hard to show how the construction goes\,\cite{GtHmonopole}. The result was of high physical interest: \emph{If you have a unified gauge theory whose covering group is compact \emph{(so that \(\pi_1\) is finite),} it allows for the construction of magnetic monopoles!} Their physical properties, such as their masses, can be calculated. In general, the monopole mass is of order \(M_W/e^2\), where \(M_W\) is the mass of the heaviest elementary vector particle in the theory.

The monopole solution of the gauge fields in a BEH system was discovered by the author and, independently, by Aleksandr Polyakov\,\cite{polymono}. He had been searching for what he called a `hedgehog' configuration. When discussing this at the Landau Institute in Moscow, it was Lev Ok\'un who noticed that the thing carries  a magnetic monopole charge (according to a footnote in his paper).

Now, if we call the QCD gauge coupling constant \(g_3\), then, if there were a BEH mechanism, there would be gauge magnetic monopoles with charges \(6\pi/g_3\), with a large mass, of order \(1/g_3^2\). But, there is no reason to assume a BEH mechanism in QCD at all. Without this mechanism, there is no reason to assume any lower limit to the energy required for gauge-magnetic charges in QCD. Better still, it is reasonable to suspect \emph{a BEH mechanism for the magnetic gauge charges} in QCD.

The effect of this would be wonderful. The QCD vortex must be the electro-magnetic dual of a Nielsen-Oleson vortex! In that case, not the magnetic monopoles but the quark charges would be confined by the vortex forces. Since triple electric monopole charges freely roam around in the QCD vacuum -- these are now simply the QCD gluons -- the quark confining forces will be active \emph{modulo} three, which is why both baryonic and mesonic states can be formed out of quarks.

Properties of the baryonic and mesonic bound states can be calculated numerically in lattice models. The QCD lattice was first discussed by Kenneth Wilson\,\cite{wilsonlattice}. Formally, one can perform the \(1/g_3\) expansion on the lattice, to observe that not only the confinement phase is realised in that formalism, but also chiral symmetry is broken exactly in the way observed, and in the way described by the qualitatively successful Gell-Mann-L\'evy sigma model.

The pieces all fell into place this way, and studying all related dynamical properties of QCD became a big industry. The QCD vacuum is still a hot topic of research. It was found to depend on a \(PC\) violating angle \(\th\) due to instantons (see below), besides the number of light fermion species. In model calculations, various types of transitions are found to take place as these numbers are varied. The QCD vacuum also gets highly non-trivial properties when put in a box in 4-space, with various possible topological boundary conditions.\,\cite{baal}

A supplementary development was the idea of \emph{jet physics}\,\cite{jets}: in high-energy scattering, jets emerge whose leading particle represents a single quark or gluon, fragmenting into mesons and baryons. The properties of these jets are calculable, so that they give us a means to compare theory with experiment. 

\newsecl{The Large $N$ limit. Planar diagrams}{planar}

Even if our understanding of QCD was greatly improving, numerical calculations continued to be laborious and voluminous. We are still looking for more efficient approximation methods, such as the highly successful loop expansion in QED. What is needed is a small parameter in the theory, in terms of which we can perform asymptotic expansions. There is one parameter that perhaps could be used for this purpose. Suppose we replace the gauge group \(SU(3)\) by a group of the form \(SU(N)\), would there be a \(1/N\) expansion? The question was asked by Claude Itzykson and Bernard Zuber\,\cite{Itz}, but they thought that nothing special happens in the infinite \(N\) limit, since diagrams with many loops will still dominate.

Yet something special does happen: of all diagrams, \emph{only the planar ones} survive\,\cite{ninfty}. Using a simple topological argument, this could be proved. A planar Feynman diagram is a diagram that can be drawn on a piece of paper without lines crossing one another. These diagrams emerge in what is now called the 't Hooft limit: \(N\ra\infty, \ \  \tl g^2\equiv g^2N\) is kept fixed. This result was very suggestive. Planar diagrams are very similar to the world sheets of strings. Could it be that confinement can be proven in this limit?

The \(N\ra\infty\) limit indeed simplifies the computation of Feynman diagrams considerably, but exactly summing all diagrams that contribute, even in this limit, is not possible with presently known techniques.  Today, we do know a lot more about this limit. It is frequently employed in ADS/CFT transformations.

\newsecl{Grand Unification}{guts}
	It was soon realised that what we call the `Standard Model' today, may well be the tip of an iceberg. Having seen that the weak force and the electromagnetic force become (partly) unified, one could ask whether unification can occur in a more drastic fashion when we look further. On the one hand, one may speculate on wilder gauge field structures at moderate energies. Direct evidence for extended gauge structures have been lacking so-far, but one is free to speculate. All sorts of models were suggested. A good attempt was the speculation that many or all of the elementary particles in today's Standard Model could turn out to be composites. The simplest construction that could lead to such a picture is a repetition of QCD at roughly a thousand times higher energies: ``Technicolor". Today's quarks and leptons are tomorrow's mesons and baryons of Technicolor. Estia Eichten and Kenneth Lane\,\cite{eichtenlane} pioneered this theory, but, as of this moment, strong supporting experimental evidence for technicolor is still lacking; many of its predictions were falsified by experiments. Multitudes of repairs were attempted, but the theory is still not in a good shape.
	
	It is very tempting to try to unify the three gauge groups seen today, \(SU(3),\ SU(2)\) and \(U(1)\), into one. Abdus Salam and Jogesh Pati, John Ellis and Dimitri Nanopoulos, together with many others, investigated various possibilities. One of the early approaches\,\cite{su5} involved the gauge group \(SU(5)\). It regards a single generation of fermions as a \(5\) plus a \(10\) representation. The sterile neutrinos form invariant singlets.
	
	Such a theory would predict the existence of physical magnetic monopoles, although these would be very heavy. It would also imply that protons are unstable, decaying into pions, electrons, positrons, muons, neutrinos or others. The decay time could be estimated. Ingenious experiments were designed and carried out, but no such decay has as yet been detected. 
	
	\(SU(5)\) is a subgroup of \(SO(10)\), and this is a nicer group, as it very elegantly puts each generation of fermionic particles (including the sterile neutrinos) in a single fermionic 16 representation\,\cite{so10}. This representation handles the fermions just as the space-time group \(SO(3,1)\) does, as if one should combine these groups into a single \(SO(13,1)\).

\newsecl{Magnetic monopoles, solitons and instantons}{mm}

The discovery of magnetic monopoles in certain unified theories demonstrated that gauge theories have a rich topological structure. It was an interesting exercise to search for more examples of this. \emph{Solitons} are particle-like configurations that are stable on a one-dimensional line. They typically describe a boundary between two equivalent vacuum configurations. These are stable if one has a double-well potential in a real scalar field variable, both potential wells being equally deep (because of a symmetry). 
Subjecting these to a rigorous quantization procedure is an interesting exercise in mathematical physics. 

The Abrikosov-Nielsen-Oleson vortex is the two-dimensional analog of that, in the sense that it can be regarded as a soliton living in two space dimensions, taking the shape of a line if we add the third space dimension. It owes its existence to the fact that the minima of the Higgs field potential form a closed loop in field space. One can lasso this loop around the line\fn{Sidney Coleman\,\cite{colemerice} gave famous lectures about this at the Erice Summerschool of Subnuclear Physics. He demonstrated the notion of topology by winding the cord of his microphone around his neck. Students got worried that he might strangle himself.}.

Next comes the magnetic monopole. It is a stable, particle-like configuration in three space dimensions. Apart from its indirect presence in QCD, no signs have been observed for its existence as a particle with a \(U(1)\) magnetic charge. Further research has shown that these particles would be special indeed. Bernard Julia and Tony Zee\,\cite{juliazee} observed that, besides their magnetic charge, monopoles can also have electric charge. This charge itself needs not obey the usual charge quantization rule. In a two-dimensional plane where we plot the allowed values of the 2-vector \((q_i,\,g_{m\,i})\), where \(q_i\) is the electric charge and \(g_{m\,i}\) the magnetic charge of  a stable monopole-like configuration \(i\), the outer products \hbox{\((q_1\,g_{m2}-q_2\,g_{m1})\)} must be integer multiples of \(2\pi\). This allows for a universal arbitrary angle: \(g_{m\,i}=\fract{2\pi}en_i\,; \ q_i=k_ie+\fract{\th}{2\pi}n_ie\), where \(n_i\) and \(k_i\) are integers.
The angle \(\th\) could be related to another angle in the unbroken theory, the instanton angle \(\th\) (see below).

An other curious feature in magnetic monopoles is due to the large value of their magnetic charges. Since elementary charged fermions carry magnetic moments with a giro-magnetic ratio of about 2, one can calculate that the dipole force between a charged fermion and a monopole (if the monopole is electrically neutral) can generate an attractive or repulsive \(1/r^2\) potential that is so strong that it may cancel the usual angular momentum term in Schr\"odinger's equation. This means that these particles reach severely modified bound state configurations, which may need extra boundary terms for the two particles coming very close. At the close positions, effects due to the underlying grand unified theory may become sizeable, even if the fermions have very low energy. One finds that \emph{baryon number decay} may become a strong force there. In short: magnetic monopoles could behave as very strong catalysts for proton decay. This discovery was made and worked out by Valery Rubakov and C.G.~Callan\,\cite{rubcal}.

There is even more to be learned from magnetic monopoles. careful analysis shows that, if we take an electric charge \(q\) orbiting a magnetic monopole with charge \(g_m\), the total angular momentum takes values \(q\,g_m/4\pi\ +\) integers. This means that a bosonic electric charge \(e\) and a bosonic magnetic monopole with magnetic charge \(2\pi/e\) will produce fermionic bound states\,\cite{jackiwrebbi}! The calculation of these states is straightforward. The fact that also the statistical properties of the bound state are those of fermions was first understood by Alfred Goldhaber\,\cite{goldh}.

Having seen soliton-like structures in 1, 2, and 3 dimensions, we can ask whether there is anything interesting going on in 4 dimensions. Are there topological structures that are stable in four dimensions? Such objects would be space-time points, or, \emph{event}-like. For that reason, they are called \emph{instantons}. The simplest classically stable instanton can be seen to occur in a scalar field theory if the self interaction is a curve with two (or more)  minima, while the vacuum state is chosen to be in the `wrong' minimum, the minimum that is locally stable but does not represent the lowest energy possible. This vacuum state would then not be absolutely stable: \emph{quantum tunnelling} could cause the vacuum to decay spontaneously into the true vacuum. The tunnelling could initiate at one spot in space and time: the instanton event. Careful analysis of the mathematical physical question how to calculate the decay probability of the false vacuum, shows that the probability is dominated by the exponent of the total action of an \(SO(4)\) invariant field configuration in Euclidean space-time, the scalar field instanton.

Non-Abelian gauge theories also have instantons. Alexander Belavin, A.\ Pol\-yakov, Albert Schwarz, and Yu. S. Tyupkin\,\cite{BPST} described a four dimensional topological solution of the classical Yang-Mills field equations in Euclidean space, calling it a four-dimensional `pseudo-particle'. They also made the interesting observation that the parity-odd expression \(\int\dd^4x\,F_{\m\n}\tl F_{\m\n}\) takes values \(32\,n\,\pi^2/g^2\) if you have \(n\) of such instantons. Now it was known that, due to the chiral anomaly, the axial current \(J_\m^A\) for massless fermions is not exactly conserved. Instead, it obeys
	\be \pa_\m J^A_\m={g^2\,L\over 16\pi^2}F_{\m\n}\tl F_{\m\n}\ ,  \eel{axialdiv}
where \(L\) is the number of flavors. This means that exactly \(2L\) units of chiral charge are annihilated or produced by this BPST instanton. Indeed, for every flavor one elementary fermion flips from left to right, or they all flip from right to left. The fact that the chirality of chiral fermions is not conserved in a BPST instanton can be understood by carefully inspecting the boundary conditions for fermions there, and noticing that these allow for a four-dimensional, zero action Jackiw Rebbi bound state.\,\cite{GtHinst}.

The BPST instanton describes tunnelling between different vacuum states in a gauge theory. These vacuum states form infinite sequences of states, each connected to one another by topologically non-trivial gauge transformations. Classically, these quantum states are disconnected; quantum mechanically, they can tunnel to one another. In the quantum tunnelling amplitude, a phase rotation  \(e^{i\th}\) may take place, where the angle \(\th\) emerges as a new constant of nature.
The intensity of this instanton-induced process can  be calculated from the exponent of the action of this instanton. The fact that there are zero energy, or action, solutions for fermions in monopoles and instantons, gives them very special physical properties. In mathematics, the numbers of such solutions are controlled by index theorems\,\cite{ahs}. The bound state solutions furthermore play a role in establishing the nature and degeneracies of the instanton solutions themselves.

In effective field models for the mesons in QCD, there had been several problems. One was the decay \(\pi^0\ra 2\g\). According to the theory of spontaneous breakdown of axial symmetry when light quarks are present, this decay should be strongly suppressed. The fact that it is not suppressed had been attributed to the Adler-Bell-Jackiw anomaly, Eq.~\eqn{axialdiv}. This was understood fairly well. However, the same symmetry should dictate that the \(\eta\) particle is the Goldstone boson of chiral symmetry, hence it should have a low mass value, comparable to the pions. The fact that the \(\eta\) mass is much bigger, presented us with the \(\eta\) \emph{problem}. How do we compute this mass? We now found that the BPST instanton is the culprit. It generates an effective action where the number of chiral charges is not conserved, such as the \(\eta\) mass term. This effective action is strong, and it also may play a role in different systems. The admixture of biquark mesons with tetra-quark mesons, found in more recent experiments, is an example of a force that appears to be induced by the same instanton\,\cite{GtHmaiani}.

These are instantons associated to the strong \(SU(3)\) interaction, whose effects are highly visible in QCD, but we also have instantons in the electro-weak \(SU(2)\otimes U(1)\) sector of the Standard Model. Being weak, these instantons carry a very \emph{large} action, and as such, the quantum tunnelling effects they describe are \emph{extremely} weak. Now it so happens that, in the electroweak theory, baryon number is unevenly spread over the left handed sector and the right handed sector of the fermion spectrum. Consequently, the instanton induces baryon number violating interactions. Each generation of fermions hands in one unit of baryon number, so that a single electroweak instanton produces a change of three units in the total baryon number: \(\D B=\pm 3\). 

In fact, one can now understand why, in the perturbative regime, the gauge anomalies have to cancel out. If they hadn't cancelled out, left helicity and right helicity particles would have been produced in such a way that electric charge is not conserved; this would be inadmissible in any gauge theory.

This instanton tunnelling effect will be exorbitantly weak, yet it is
thought perhaps to play a role in the genesis of baryon- and lepton asymmetry in the early universe\,\cite{baryogenesis}. This is because, at very high temperatures, the transition can occur classically, without tunnelling. The transition goes \emph{via} a classical, unstable field configuration called \emph{sphaleron}, pioneered by Frans Klinkhamer and Nick Manton\,\cite{klinkh}. 

Fact is, that instantons induce the kind of \(CP\) violation needed to understand the matter-antimatter asymmetry in the universe. Quantitative calculations however, show that there is still a problem: the matter-antimatter imbalance in our universe seems to be greater than one would expect from these calculations.
	
\newsecl{Supersymmetry and gravity}{sugra} 

While all these developments took place, and many new insights developed, there were numerous other, related researches that greatly affected our understanding of quantized fields. Without the fantastic experimental efforts over almost a century, we would not have been able to answer any of our questions, or even ask them. Advances in numerical techniques, such as lattice theories using supercomputers, enabled us to investigate the QCD vacuum structure. 

But a more peculiar feature turned out to yield all sorts of information that shines a new light on many of our findings: supersymmetry. It started when various coincidences were discovered in models that related fermionic and bosonic fields and their properties. It seemed more and more obvious that fermions and bosons, somehow, are related. In the early 1970's, Julius Wess and Bruno Zumino wrote down their first models that exhibited supersymmetry, first for lower-dimensional models, then for chiral theories with scalars and chiral Dirac fermions, later for gauge theories with fermionic and scalar fields, which displayed higher forms of supersymmetry, called \(\NN=2\) and \(\NN=4\) supersymmetry, and when Sergio Ferrara and Peter van Nieuwenhuizen hit upon some peculiar coincidences in perturbative quantum gravity, also the gravitational field was found to have supersymmetric connections with other fields. Here, we could continue all the way to \(\NN=8\) supergravity.

Supersymmetric versions of QCD gave new insights in the various confinement modes. Furthermore, these theories appeared to display rich duality structures, relating one model to a different one. Supersymmetry also turned out to serve as an essential ingredient of the quantized relativistic string theories, originally designed to help us understand quark confinement, but later seen to function even better when used as a super unifying theory, connecting the gravitational force to all other forces in nature. Without supersymmetry, these systems cannot be quantized in a completely satisfactory manner, so that this subject is now known as \emph{superstring theory}. This theory is being advocated as an extremely promising approach towards quantising gravity, and it has indeed deeply transformed the topic of reconciling General Relativity with Quantum Mechanics. To what extent superstring theories have become more than promising, is discussed in other chapters of this book. 

There is one thing theoreticians do not seem to be able to do. Being so ambitious as to attempt to produce a \emph{Theory of Everything}, did not help us answering in a satisfactory manner a very basic question: \emph{Why is our universe so complex?} What we mean here is the obvious observation that our world is controlled by large and small numbers. There are numerous examples of this. Not only is Sommerfeld's fine-structure constant \(\a\) fairly small, \(\a=1/137.036\), and \(m_p/m_e\) is fairly large, \(\approx 1836.15\), while the mass ratio for the electron and its neutrino is somewhere in the range \(10^5\) to \(10^8\), but we have much more extreme numbers:
	\be	{M_{\mathrm{Planck}}\over m_{\mathrm{proton}}}=1.301\times 10^{19}\ ;\qquad \L\,L_{\mathrm{Planck}}^2\approx 6\times 10^{-122}\ .	\eel{ratios}
The universe is so big and complex because its laws are based on very large and very small dimensionless numbers. Where do these numbers come from? This is the \emph{hierarchy problem}. The only answer given today is the {anthropic principle}: \emph{these numbers have the values they have because these would be the conditions for having life in the universe.} We wouldn't have been there to ask the question, if these numbers had different values.

This argument would have been acceptable if the theory would have given us a list of discrete numbers from which we could have chosen the constants. No theory today can give us such a list.

	\newsecl{Calculations}{calc}
	
In theoretical physics, models are often referred to as theories (Yang-Mills theory, String theory, \dots) while theories are often called models. The Standard Model is a case in point. It now appears to describe the world of the fundamental particles so well, that it has become much more than a model. Time has come to refer to it as the \emph{Standard Theory}, as in the title of this book.

We have learned how to do the perturbation expansion for this theory to all orders (in principle), we identified various non perturbative features such as soliton-like solutions, confinement mechanisms, tunnelling through instantons, and effective descriptions of confined particles such as quarks and gluons in terms of \emph{jets}. Now the question is how to do calculations more efficiently and accurately. 

The perturbation expansion is formally divergent, but when the expansion parameter is not too big, the first two or three terms give already quite accurate results, and sometimes, such as in the case of  \(g-2\) for the electron or muon, impressive precision can be reached by including even higher terms.
In many cases, however, this expansion is not good enough. Instanton effects are fundamentally non-perturbative, being dominated by exponential terms such as \(e^{-8\pi^2/g^2}\), but subsequently,  these can, in principle, be supplemented by ordinary perturbation expansions. A persistent feature is that individual Feynman diagrams yield logarithms of mass / momentum ratios that can became rather large. One is then interested in identifying these `leading logarithms' and to resum them. The leading logarithms can be understood by renormalization group arguments, and resumming them, in particular for strongly interacting particles such as quarks and gluons, has become an important industry. 

An equally challenging question is how the perturbation expansion may behave at very high orders, so as to understand how their contributions can be rearranged and combined. Here again, instantons may play a role. Combining instantons and anti-instanton in pairs, such that these have vanishing topological winding numbers,  their contribution is found to be \(\OO(e^{-16\pi^2/g^2})\), and this indicates that the perturbation series may diverge as
	\be C\sum_{n=0}^\infty \,n!\,g^{2n}/{(16\pi^2)}^n\ , \ee
a kind of behaviour that can be understood by performing Borel resummation methods. Indeed, as was argued by Lev Lipatov\,\cite{lipatov}, it is well-known from ordinary quantum mechanics, that tunnelling phenomena cause perturbation expansions to diverge this way. In terms of Feynman diagrams, one can observe that this divergence is related to the fact that the total number of Feynman diagrams diverges as \(A^nn!\) at \(n^{\mathrm{th}}\) order, for some number \(A\).

There is, however, an other source of divergent behaviour in perturbation expansions. This has nothing to do with tunnelling, but is due to renormalization. If we consider a self-energy diagram that requires renormalization, it develops a logarithmic dependence of the momentum \(k^2\). When a propagator with such a self-energy correction built in, in turn occurs inside a loop diagram, one finds that integration over the momentum \(k\) generates \(n!\)-dependence because of that logarithm. Now this observation can be generalized by including renormalization group arguments to understand the \(k\)-dependence of sub-diagrams, and so get a more systematic understanding of \(n!\) divergences of higher loop diagrams. The numbers of such diagrams increases geometrically with higher orders, that is, with powers \(A^n\) but without \(n!\). The \(n!\) dependence now comes from renormalization. Because of the similarity between these effects and the instanton effects, we attribute the new \(n!\) dependence to \emph{renormalons}. Should the contributions of renormalons and instantons be computed separately and just be added into the final expressions? No, by carefully studying the \emph{instanton - anti-instanton gas}, and the way the instantons here behave under scaling, it was found that the instanton and renormalon effects blur in a more interesting way\,\cite{shuriak}.

Equipped with a better understanding of the perturbation expansion, could we now ask the question whether we can really understand quantum field theory beyond this expansion? Are there instances where the perturbation expansion can be made absolutely convergent? Should we not consider the axiomatic construction of all quantum field theory amplitudes, relying exclusively on convergent summations, as a mandatory step in developing true understanding of nature?

Attempts in that direction have been made. If, due to renormalization group behaviour, the coupling strengths of a theory run to infinity at high energies, then there is little chance that the theory can be rigorously defined, because the Landau ghost simply destroys it. Such theories can at best be \emph{effective} theories, valid as approximative models only at energies well below the Landau ghost. Beyond the Landau ghost, the couplings of the original theory would run in the wrong direction, showing that unitarity, locality, and/or causality have broken down. This means that the theory in that domain would have to be replaced by something altogether different. In asymptotically free theories, there is much more reason for hope that they can be rigorously constructed, but even here, the mathematical proof has not been given\fn{In fact, the question how to give a mathematically sound formulation of such theories, in particular proving that they have a mass gap, is one of the ``Millennium Problems" formulated by the Clay Mathematics Institute, Providence, USA.}.

What could be done is that we proved the existence of a planar limit, or the limit \(N\ra\infty\), with \(g^2N\) fixed, for an asymptotically free theory, when both the far ultraviolet and the far infrared regions are controlled by the ordinary coupling constant expansion. From a physical point of view, such theories are utterly uninteresting, but, here at least, we could prove their mathematical existence\,\cite{GtHninftyconstr}.

	{ \footnotesize

\end{document}